\documentclass[journal=jpclcd,manuscript=letter, layout=twocolumn]{achemso}

\usepackage[version=3]{mhchem}
\usepackage{graphicx, color, soul}
\graphicspath{{Figures/}}
\AbstractOn

\newcommand{\si}[1]{\text{Si}_{#1}}

\newcommand{\sinom}{\text{Si}_{n}\text{O}_{m}}
\newcommand{\cl}[2]{\text{Si}_{#1}\text{O}_{#2}}
\newcommand{\siot}{\text{SiO}_{2}}
\newcommand{\sio}{\text{SiO}}

\renewcommand{\vec}[1]{\mathbf{#1}}

\author{S. V. Lepeshkin}
 \affiliation [Skoltech] {Skolkovo Institute of Science and Technology, Skolkovo Innovation Center, Nobel St. 3, Moscow 143026, Russia}
 \alsoaffiliation [LPI] {P.N. Lebedev Physical Institute of the Russian Academy of Sciences - 119991 Leninskii prosp. 53, Moscow, Russia}
 \email{S.Lepeshkin@skoltech.ru}
\author{V. S. Baturin}
 \affiliation [Skoltech] {Skolkovo Institute of Science and Technology, Skolkovo Innovation Center, Nobel St. 3, Moscow 143026, Russia}
 \alsoaffiliation [LPI] {P.N. Lebedev Physical Institute of the Russian Academy of Sciences - 119991 Leninskii prosp. 53, Moscow, Russia}
\author{Yu. A. Uspenskii}
 \affiliation [LPI] {P.N. Lebedev Physical Institute of the Russian Academy of Sciences - 119991 Leninskii prosp. 53, Moscow, Russia}
\author{Artem~R.~Oganov}
 \affiliation [Skoltech] {Skolkovo Institute of Science and Technology, Skolkovo Innovation Center, Nobel St. 3, Moscow 143026, Russia}
 \alsoaffiliation [MIPT] {Moscow Institute of Physics and Technology, Dolgoprudny, Moscow Region 141700, Russia}
 \alsoaffiliation [NPU] {Northwestern Polytechnical University, Xi'an, Shaanxi 710072, People's Republic of China}

\title{Simultaneous prediction of atomic structure and stability of nanoclusters in a wide area of compositions}

\begin{document}

\begin{tocentry}
\center
\includegraphics[height=5cm]{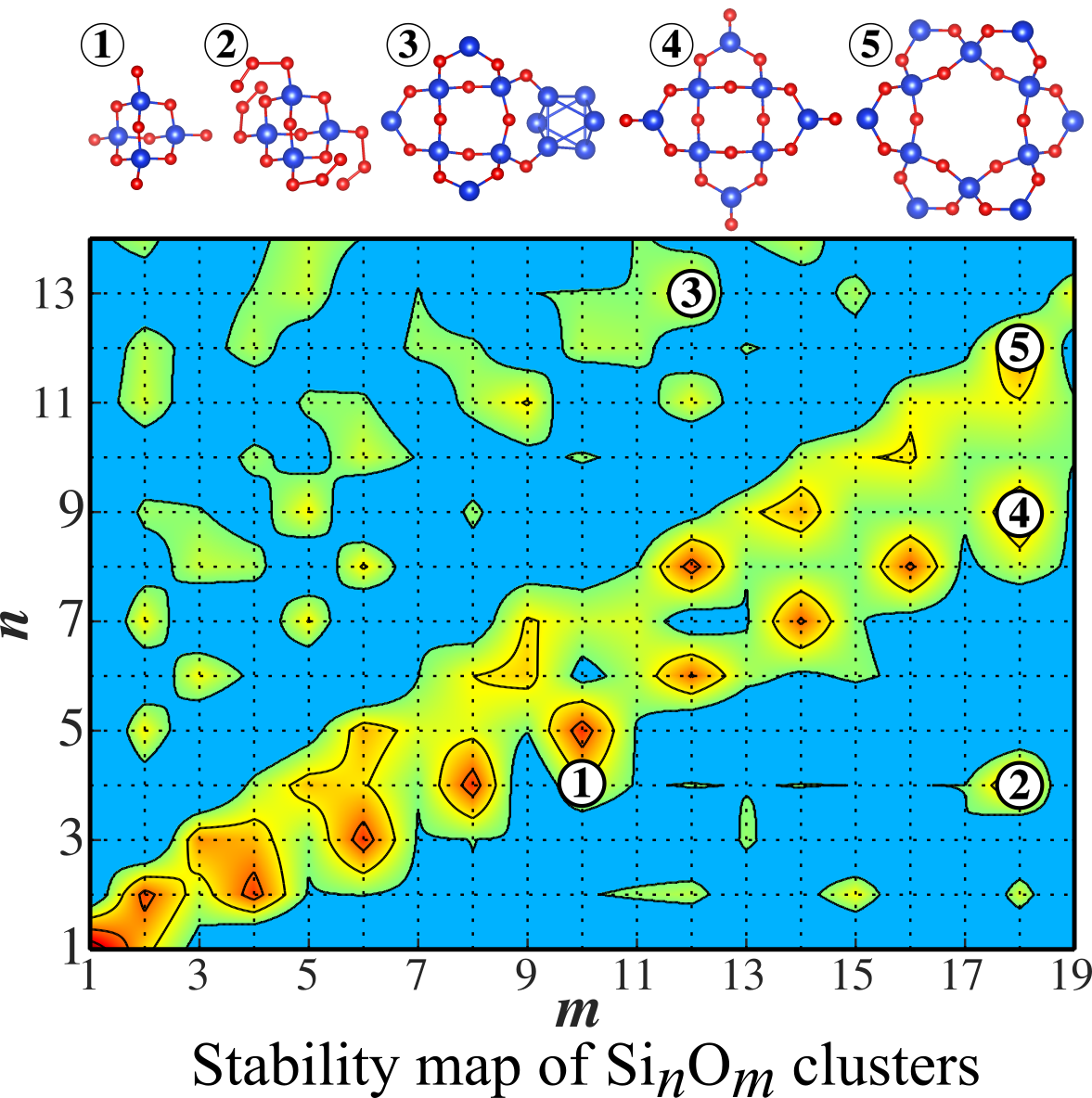}
\label{fig:For Table of Contents Only}
\end{tocentry}






\begin{abstract}
We present a universal method for the large-scale prediction of the atomic structure of clusters. Our algorithm performs the joint evolutionary search for all clusters in a given area of the compositional space and takes advantage of structural similarities frequently observed in clusters of close compositions. The resulting speedup is up to 50 times compared to current methods. This enables the first-principles studies of multi-component clusters with full coverage of a wide range of compositions. As an example, we report an unprecedented first-principles global optimization of 315 $\sinom$ clusters with $n \leq 15$ and $m \leq 20$. The obtained map of Si-O cluster stability shows the existence of both expected $(\siot)_n$ and unexpected (e.g. $\cl{4}{18}$) stable (“magic”) clusters, which can be important for miscellaneous applications.
\end{abstract}

The unique properties of nanoparticles are extensively used in optoelectronics, photovoltaics, photocatalysis, biomedicine, etc. These properties are closely linked to the atomic structure of particles, what is more explicit in the small particles and nanoclusters \cite{Alivisatos1996, Petkov2013}. Despite the importance of knowing the structure, its experimental determination remains very difficult \cite{Foster2018}. For this reason the main body of structural information on clusters is obtained via first-principles calculations \cite{Catlow2010} which were mostly done either for monoatomic clusters or for binary clusters of stoichiometric composition corresponding to the bulk compounds, while clusters of general composition were studied only in few publications\cite{Bhattacharya2013, Lepeshkin2016}. Such an accent in \textit{ab initio} research ignores the fact that the chemistry of clusters is much richer than that of solids because of a large share of surface atoms. Multi-component clusters often have stable compositions, which are far from chemical compounds presented in the bulk $x$--$T$ phase diagram. This is of interest not only for basic chemistry of clusters. It significantly increases the scope of candidate nanomaterials for practical applications such as: the development of efficient and affordable catalysts \cite{Petkov2018, Petkov2018nanoen} and magnets \cite{Celika}, the investigation of complex processes of nucleation and particle growth \cite{Bromley2016, Wang1998, Zhang2003}, etc.

The bottle-neck of first-principles activity in cluster studies is the computational cost of atomic structure determination, which is a global optimization of the total energy among all possible atomic configurations. There are several methods of structure prediction (basin and minima hopping \cite{Wales1997, Goedecker2004}, simulated annealing \cite{Deem1987}, evolutionary algorithm \cite{Oganov2006}, etc.), however they all involve thousands of local optimizations (relaxations) even for finding a structure of one cluster. In the applications mentioned above, the computation of atomic structure and the screening of stability and properties are required in wide regions including hundreds of cluster compositions, therefore such first-principles investigations turn out extremely exhausting. To reduce the computational cost, the global optimization is frequently performed in combination with semiempirical methods of force fields \cite{Li2009, Rehman2011}. The success greatly depends on the model potential, which is often difficult to make sufficiently accurate. Here we suggest a different approach to this problem, which does not invoke semiempirical potentials at all. Our method predicts all clusters in the whole given area of compositions simultaneously in a highly efficient manner that incorporates exchange of structural information, i.e. learning between clusters of different compositions. The effectiveness of our approach is based on the frequent similarity of structural motifs in clusters of close compositions. We will refer our technique to as variable-composition cluster search in contrary to the previous, fixed-composition approaches.

Our method is derived from the evolutionary algorithm implemented in the USPEX code\cite{Oganov2006, Oganov2011, Lyakhov2013}, proved to be successful for predicting novel materials\cite{Kvashnin2018, Semenok2018}. Briefly, the algorithm is based on the analogy with natural selection: first population of structures (the 1-st generation) is created randomly and locally optimized. Certain percentage of best structures serves as parents for the next generation, which is produced using the so-called variation operators: cross-over, mutation, etc. Then the next population is locally optimized and so on, until the convergence is achieved.

Our technique performs a joint evolutionary search for the whole given area of compositions at once. It required introducing two major innovations. First, the selection of the fittest configurations (selecting best parents) is performed on equal footing for clusters of all compositions. Second, we developed new variation operators to provide the transfer of structural information between clusters of different compositions.

Let’s consider the new selection procedure in more detail. It is grounded in the notion of ``magic'' clusters. We classify a cluster as magic if a pair of such clusters is stable against the transfer of one atom between them. Giving it a formal description, we denote a composition of cluster with the formula X$_{n_1}$Y$_{n_2}$...Z$_{n_k}$ by a vector $\vec{n}=(n_1,\,n_2,\ldots,\,n_k)$. If an atom of sort $s$ is added or removed, the composition becomes $\vec{n}_s^\pm=(n_1,\,\ldots,\,n_s\pm1,\ldots,\,n_k)$. The cluster is magic if the second-order differences of the energy:
\begin{equation}
\label{eq_1}
    \Delta_{ss}E(\vec{n}) = E(\vec{n}_s^+)+E(\vec{n}_s^-)-2E(\vec{n})
\end{equation}
are positive for all sorts $s$. Having the (non-regular) set of magic compositions, we build the reference energy surface $E_\mathrm{ref}(\vec{n})$ as a piecewise linear interpolation of 'magic' energies over all given area of compositions. A configuration is classified as 'best' and selected to participate as a parent for the next generation, if its energy falls in the interval $E_\mathrm{conf}(\vec{n})-E_\mathrm{ref}(\vec{n})\leq\Delta E_\mathrm{sel}$. The interval $\Delta E_\mathrm{sel}$ is defined so that the share of the best configurations is $N^\mathrm{best}/N \sim 0.6-0.8$. We note that the global thermodynamic stability of clusters is not required, as any cluster system is unstable with respect to growth or coalescence: the only truly thermodynamically stable cluster is the infinite one, i.e. a crystal. This situation is totally different from the variable-stoichiometry prediction of crystalline structures \cite{Oganov2010}, which provides only the thermodynamically stable phases.

Our method uses all variation operators of the standard fixed-composition approach\cite{Lyakhov2013}: (1) creating structures with random point symmetry; (2) permutations of chemically different atoms; (3) softmutation (displacement of atoms along eigenvectors of the softest vibrational modes) and (4) heredity (creating child structure from fragments of two parents). Among these operators only heredity is suitable for structural exchange between clusters of different compositions and was modified accordingly. To further enhance such exchange, three new variation operators are introduced: (5) transmutation (change of chemical identity of randomly selected atoms), (6) removal of one atom from the cluster, and (7) addition of one atom to the cluster.

For operators (6) and (7) a location, where one atom should be removed or added, is of importance. The choice of an atom $i$ to be removed is defined by its effective coordination number $O_i$:
\begin{equation}
    \label{eq_2}
    O_i = \frac{\sum_j\exp(-(r_{ij}-R_i-R_j)/d)}{\max_j\exp(-(r_{ij}-R_i-R_j)/d)}
\end{equation}
Here $r_{ij}$ is the distance between atoms $i$ and $j$, $R_i$ and $R_j$ are the covalent radii of atoms $i$ and $j$ depending only on their chemical identities, and $d=0.23\,\text{\AA}$ is the empirically determined parameter\cite{Cordero2008}. The $i$-th atom is removed with a probability $p_i$ proportional to $\max_{i\in s}[O_i]-O_i$, where $s$ is the identity of an atom $i$. Thus, the removal of weakly-bound atoms is preferable. The same applies to the choice of an atom to which an additional atom should be attached. Such atom addition enhances coupling between a weakly bound atom and the remainder of cluster that gives the maximum gain in binding energy. 

We note that new add/remove atom operators usually do not create new structural motifs, but spread the good ones between different compositions, thus providing the thorough exploration of the low-energy areas of the landscape. Other operators, considering their greater stochasticity, are responsible for a sufficient level of structural diversity. Such an approach provides a balance between scattering of trials for the effective sampling of search space and focusing in the most promising regions which is crucial for any heuristic optimization scheme.

The seven variation operators show varying efficiency for different systems and even at different stages of evolutionary search. To take this into account we developed a control procedure which changes on-the-fly the number of offspring produced by each operator considering their efficiency at previous stages of the search \cite{Bushlanov2018}. To increase structure diversity, our method uses yet another, very powerful tool, namely, “antiseeds”. It gives energy penalty for structures which are best for too many generations\cite{Lyakhov2013}. This expedient allows one to reliably determine structures of all non-magic clusters belonging to the area of compositions, as well as low-energy isomer structures. 

One may recollect that many researchers tried to exploit structural similarity of clusters of close compositions “by hand”\cite{Bromley2016, Lu2003, Wang2008, Liu2016}. However, the number of possible configurations is still extremely large. For this reason such manual sampling requires too much effort and often fails. In contrast, our method makes similar sampling fully automatically, integrating it into the general evolutionary process. Combined functioning of all variation operators, new selection scheme, “antiseed” technique and other features give synergistic effect that results in the high performance of our method.

\begin{figure}[t!]
      \includegraphics[width=8cm]{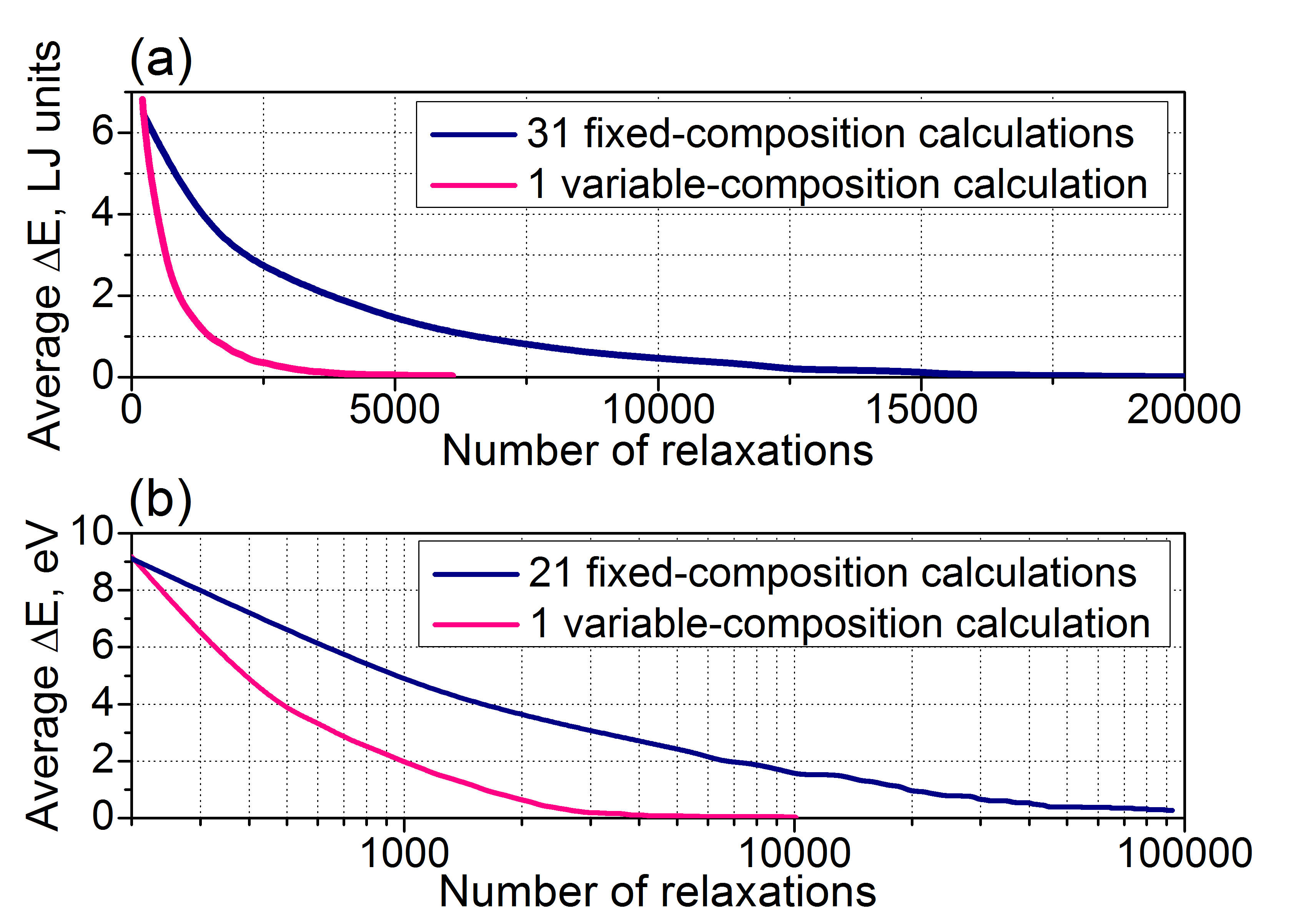}
      \caption{Convergence rate of variable- and fixed-composition methods. Energy deviations from the ground states averaged over all compositions as a function of number of relaxations is given for (a) Lennard-Jones clusters with 30-60 atoms and (b) $\sinom$ clusters ($n=6-8$ and $m=10-16$) calculated by the MNDO method (the logarithmic scale is used).}
      \label{fig:tests}
\end{figure}

We test our technique on two model systems: Lennard-Jones clusters with 30-60 atoms and $\sinom$ clusters ($n=6-8$ and $m=10-16$) within the semiempirical MNDO approach as implemented in the MOPAC package\cite{MOPAC2016}. We compare the convergence speed of the new approach with the standard fixed-composition technique implemented in the USPEX code\cite{Lyakhov2013}. Fig. \ref{fig:tests} shows the energy deviations from the ground state averaged over all compositions as a function of the number of relaxations for both methods. This integral characteristic of the convergence rate shows a great speedup of the new method ($\sim 5$ times for LJ clusters and up to 50 times for $\sinom$ clusters). A more thorough analysis presented in Supplementary Information (SI) shows high efficiency of new variation operators which produce most of low-energy offspring of the new method.

As the first real application, we chose $\sinom$ clusters owing to their practical importance and great structural diversity (crystalline silica alone has 14 structural forms). The search was performed at the \textit{ab initio} level within the unprecedentedly large range of compositions ($1\leq n \leq 15$ and $0\leq m \leq 20$ i.e. 315 cluster compositions). We note that the earlier investigations of $\sinom$ were either done for relatively small clusters ($n \leq 7$) \cite{Lepeshkin2016, Lu2003, Liu2016, Zang2006, Caputo2009} or focused on stoichiometric compounds $(\sio)_n$ or $(\siot)_n$\cite{Catlow2010, Bromley2016, Wang2008, Hu2010, Reber2008}. Even for these compounds it is often seen that newer papers report lower-energy structures than the older ones. As an example, we mention recent study of $(\sio)_n$ clusters which were constructed ``by hand'', joining fragments of $(\siot)_n$ and $\si{n}$ clusters together \cite{Bromley2016}. Surprisingly, these structures turned out to be better than those reported in all previous studies using global optimization techniques.

\begin{figure*}[t!]
\center
      \includegraphics[width=16.5cm]{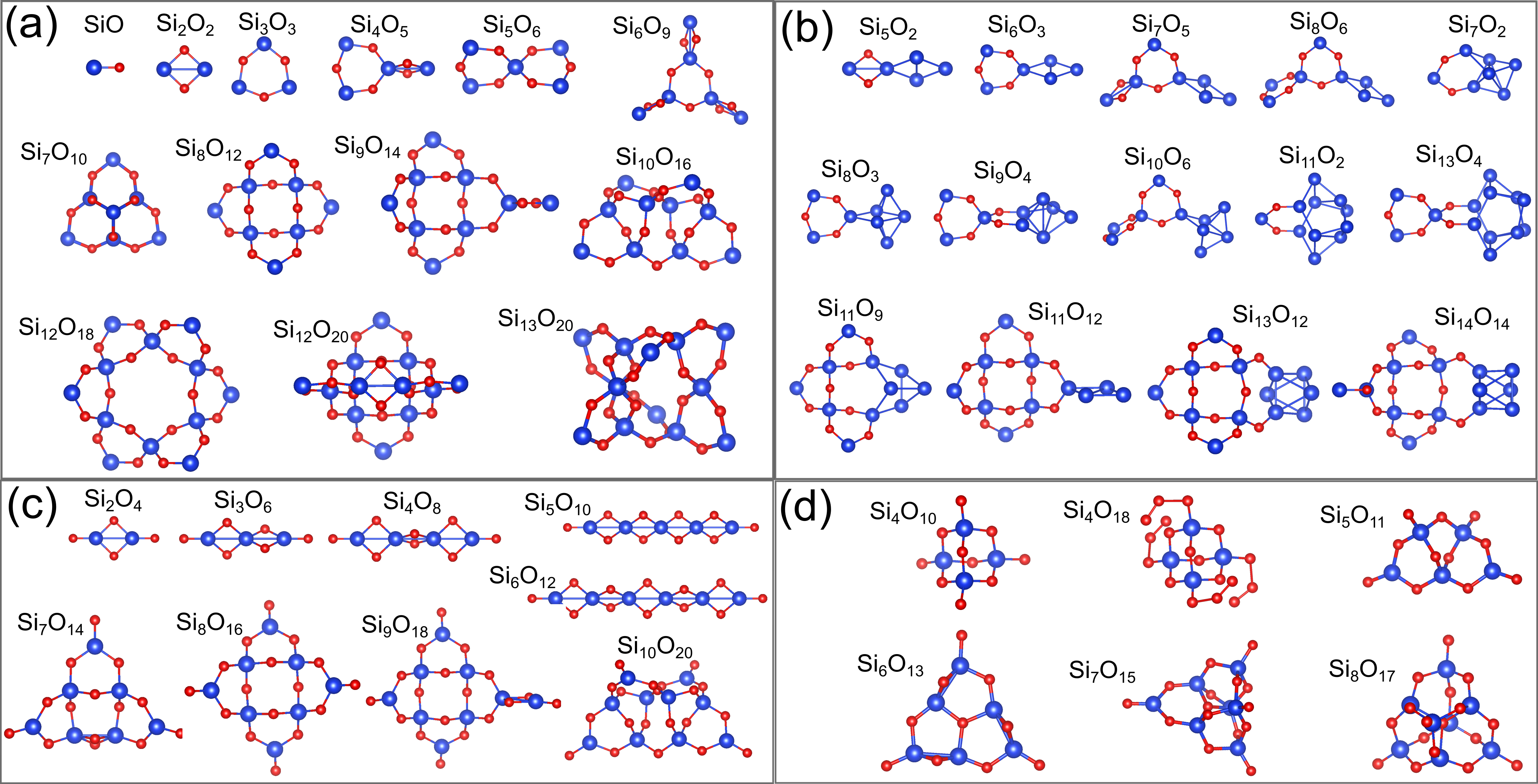}
      \caption{Optimal structure of selected stable $\sinom$ clusters, divided into 4 groups: (a), (c) the most stable clusters with $n:m\sim2:3$ and $n:m=1:2$ respectively; (b) silicon-rich clusters and (d) super-oxidized clusters ($m\geq2n+1$)}
      \label{fig:struct}
\end{figure*}

We perform our global optimization combined with density functional calculations within the PAW-PBE approximation implemented in the VASP code \cite{Kresse1993, Kresse1996}. The energies of 10 best structures for each composition were refined using the GAUSSIAN code \cite{Frisch2009} with the B3LYP/6-311+G(2d,p) approach\cite{Stephens1994}. Comparison of our results with earlier  publications\cite{Catlow2010, Lepeshkin2016, Bromley2016, Lu2003, Wang2008, Liu2016, Zang2006, Caputo2009, Hu2010, Reber2008} showed that 101 optimal structures of $\sinom$ clusters were correctly reproduced by us, 17 better ones were found, and 197 clusters were studied for the first time to our best knowledge (see table 1 in SI). Fig. \ref{fig:struct} shows $\sinom$ stable (magic) clusters which are divided into four groups discussed below. The optimal structures of all 315 clusters are given in SI (table 2).

Scanning over wide composition areas reveals trends in cluster structure and related properties. To illustrate this point, we explore the stability patterns of $\sinom$ clusters using two criteria. The first one characterizes the resistance toward the transfer of Si or O atoms between clusters. It calculates second-order differences over Si and O atoms ($\Delta E_{nn}(n,m)$ and $\Delta E_{mm} (n,m)$ of Eq. \eqref{eq_1}) and takes the minimal one: $\Delta_\mathrm{min}(n,m) = \min\{\Delta_{nn}E(n,m),\Delta_{mm}E(n,m)\}$. The second criterion calculates dissociation energies for all possible fragmentation channels $\sinom \rightarrow \cl{k}{l}+\cl{n-k}{m-l}$ with $0 \leq k \leq n$ and $0 \leq l \leq m$:

\begin{equation}
\label{eq_ediss}
    E_\mathrm{diss}(n,m,k,l)=E(k,l)+E(n-k,m-l)-E(n,m)
\end{equation}
and picks the lowest of them: $E_\mathrm{diss}(n,m)=\min_{k,l}\{E_\mathrm{diss}(n,m,k,l)\}$.The higher is $E_\mathrm{diss}(n,m)$, the more resistant to fragmentation the cluster is. In stable clusters both $\Delta_\mathrm{min}(n,m)$ and $E_\mathrm{diss}(n,m)$ should be positive, while a negative value of $\Delta_\mathrm{min}(n,m)$ or $E_\mathrm{diss}(n,m)$ is a sign of instability. 

\begin{figure*}[t!]
\center
      \includegraphics[width=16.5cm]{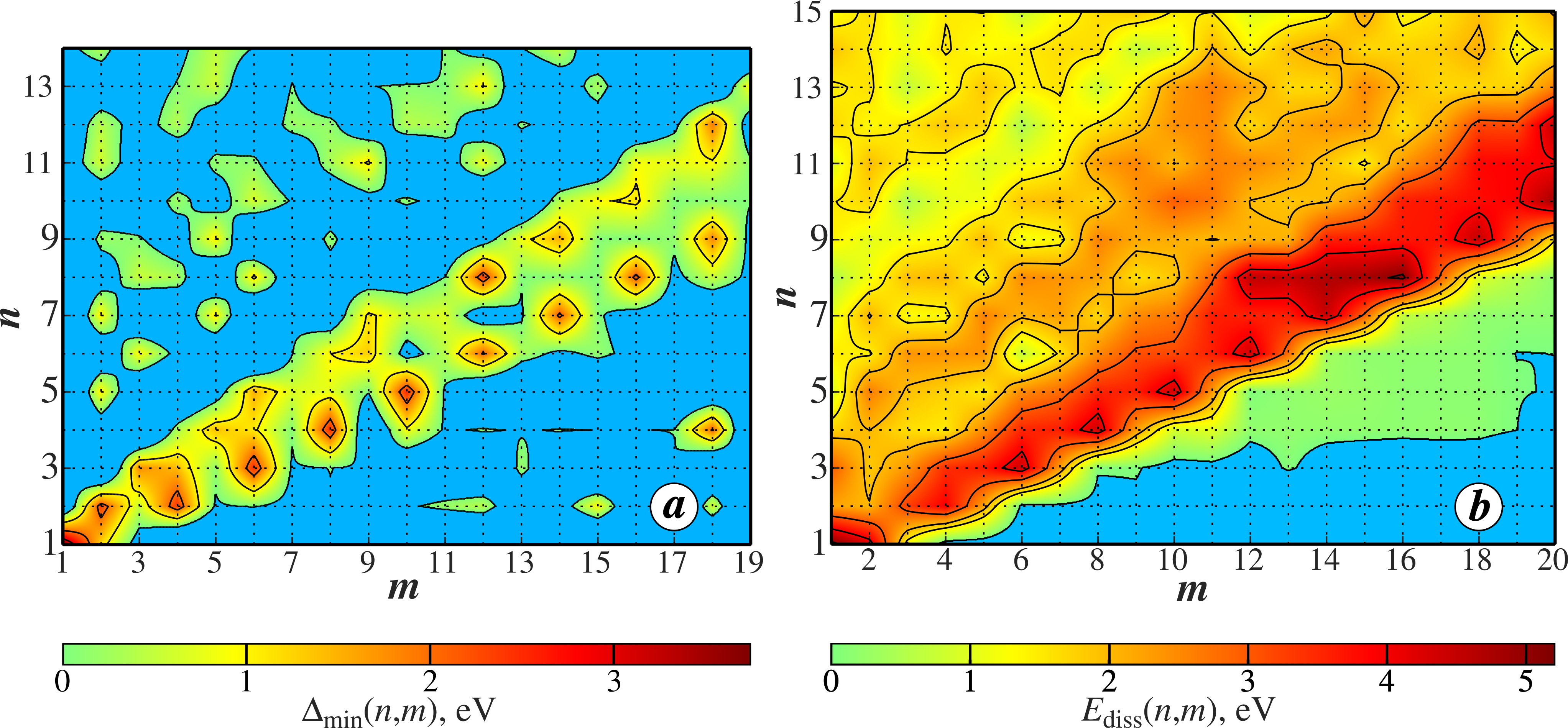}
      \caption{The relief map of stability in $\sinom$ clusters according to two criteria: (a) $\Delta_\mathrm{min}(n,m)$ (in eV) as a function of n and m, showing resistance to transfer of Si or O atom between two identical clusters, (b) $E_\mathrm{diss}(n,m)$ showing resistance to dissociation into fragments. Regions of instability are marked by blue color}
      \label{fig:maps}
\end{figure*}

Fig. \ref{fig:maps} shows the contour maps of calculated $\Delta_\mathrm{min}(n,m)$ and $E_\mathrm{diss}(n,m)$ as functions of $n$ and $m$. In both figures, the areas of high stability look like mountain ridges or islands. As expected, silica $(\siot)_n$ clusters are highly stable according to both criteria. Surprisingly, $\sinom$ clusters with $n\sim 2/3m$ exhibit comparable stability. These clusters resemble $(\siot)_n$, but are constructed of Si--O--Si bridges only and have no Si=O double bonds (see Fig. \ref{fig:struct}a). As seen in Fig. \ref{fig:maps}a there are also several minor stability islands running along $n\sim4+2m/3$, $n\sim 6+2m/3$ and $n\sim 10+2m/3$. Such non-stoichiometric compounds are also rather stable according to the second criterion (Fig. \ref{fig:maps}b). They contain excessive silicon, which segregates as a compact group of Si atoms, attached to the skeleton of Si--O bonds only (see Fig. \ref{fig:struct}b). Such clusters are of interest due to experiments on growth of long silicon nanowires from gas-phase SiO [24]. 
Another interesting class of stable compounds is super-oxidized $\cl{4}{10}$ and $\cl{n}{2n+1}$ ($n\geq 5$) clusters (see Fig. \ref{fig:struct}d). The latter have relatively low values of $\Delta_\mathrm{min}$, but quite high values (2.5-3 eV) of $E_\mathrm{diss}(n,m)$. We also distinguish the \ce{Si4O18} cluster which is notably stable by the first criterion. Interestingly, this cluster is similar to recently synthesized \ce{P4O18} cluster\cite{Dimitrov2003}, but has free-ending O$_3$ groups instead of closed ones in \ce{P4O18} (see Fig. \ref{fig:struct}d). The important feature of super-oxidized clusters is spin-polarized groups with O-O bonds providing their high reactivity, which may determine the toxicity of silica particles\cite{Lepeshkin2016, Baturin2018}. 

The exploration of $(\sio)_n$ clusters is of interest for astrophysics due to the presence of SiO molecules in the circumstellar space and their role in formation of silicates. We note that these clusters do not form a distinct range of $\Delta_\mathrm{min}$ or $E_\mathrm{diss}$. For this reason they can transform to neighboring, more stable clusters that should be taken into account in interpreting their optical signatures.

Concluding, we have developed a new method for simultaneous prediction of structures of clusters in vast areas of composition. Comparing to currently used methods our approach demonstrated 5-50 times speed-up, allowing for massive ab-initio calculations of nanoclusters at reasonable costs. The availability of such an efficient tool opens the door to wide exploration of trends in chemistry of multi-component nanoclusters, to study cluster features connected with the bulk $x$--$T$ phase diagram, and to the search for new, nonstoichiometric 'islands of stability', which can be interesting for applications. These prospects for nanomaterials science are supported by our first-principles study of $\sinom$ clusters in a very wide range of compositions. We present the overall picture of stability in these clusters and show numerous ridges and islands of stability, which are very distinct from well-studied silica clusters. We hope, this first attempt gives strong impetus to wide \textit{ab initio} research in the plethora of important multi-component cluster systems with rich chemistry.

\begin{acknowledgement}
A.R.O. is supported by the Russian Science Foundation (No. 16-13-10459). We also thank Russian Foundation for Basic Research (No 16-02-00612, 18-32-00991 and 17-02-00725). Calculations were performed on the Rurik supercomputer at MIPT, on MVS-10p cluster at the Joint Supercomputer Center (Russian Academy of Sciences, Moscow, Russia) and Lobachevsky cluster at the University of Nizhny Novgorod.
\end{acknowledgement}

\end{document}